\documentclass[sigconf]{acmart}

\usepackage{natbib}
\usepackage{algorithmic}%
\usepackage{mathtools}
\usepackage[skip=2pt]{caption}
\usepackage{xspace}
\usepackage{subcaption}

\newcommand{\scinobo}{\textsc{SciNoBo}\xspace}
\newcommand{\modelsresults}{\textsc{models}\xspace}
\newcommand{\dann}{\textsc{dann}\xspace}
\newcommand{\danntopone}{\textsc{dann-top-1}\xspace}
\newcommand{\danntoptwo}{\textsc{dann-top-2}\xspace}
\newcommand{\scinobocitreftopone}{\textsc{\textsc{S}ci\textsc{N}o\textsc{B}o-citref-top-1}\xspace}
\newcommand{\scinobocitreftoptwo}{\textsc{\textsc{S}ci\textsc{N}o\textsc{B}o-citref-top-2}\xspace}
\newcommand{\topone}{\textsc{top-1}\xspace}
\newcommand{\plosone}{\textsc{plos one}\xspace}
\newcommand{\toptwo}{\textsc{top-2}\xspace}
\newcommand{\scinobocitref}{\textsc{\textsc{S}ci\textsc{N}o\textsc{B}o-citref}\xspace}
\newcommand{\scinoboreftopone}{\textsc{\textsc{S}ci\textsc{N}o\textsc{B}o-ref-top-1}\xspace}
\newcommand{\scinoboreftoptwo}{\textsc{SciNoBo-ref-top-2}\xspace}
\newcommand{\scinoboref}{\textsc{SciNoBo-ref}\xspace}
\newcommand{\scinobopubtopone}{\textsc{SciNoBo-pub-top-1}\xspace}
\newcommand{\scinobopub}{\textsc{SciNoBo-pub}\xspace}
\newcommand{\scinobopubtoptwo}{\textsc{SciNoBo-pub-top-2}\xspace}
\newcommand{\acmag}{\textsc{mag}\xspace}
\newcommand{\crossref}{\textsc{crossref}\xspace}
\newcommand{\tfidf}{\textsc{tf-idf}\xspace}
\newcommand{\selfattention}{\textsc{self-attention}\xspace}
\newcommand{\rnn}{\textsc{rnn}\xspace}
\newcommand{\fasttext}{\textsc{fasttext}\xspace}
\newcommand{\adam}{\textsc{adam}\xspace}
\newcommand{\sciencemetrix}{\textsc{sciencemetrix}\xspace}
\newcommand{\macrofone}{\textsc{macro-f1}\xspace}
\newcommand{\microfone}{\textsc{micro-f1}\xspace}
\newcommand{\mylabel}{\textsc{label}\xspace}
\newcommand{\doi}{\textsc{doi}\xspace}
\newcommand{\mytitle}{\textsc{title}\xspace}
\newcommand{\myabstract}{\textsc{abstract}\xspace}
\newcommand{\publishedvenue}{\textsc{published venue}\xspace}
\newcommand{\weightedmacrofone}{\textsc{weighted-macro-f1}\xspace}
\newcommand{\emnlp}{\textsc{emnlp}\xspace}

\usepackage{url}
\usepackage{graphicx}
\usepackage{amsmath}

\usepackage{longtable}
\usepackage{multirow}
\usepackage{enumitem}% http://ctan.org/pkg/enumitem

\AtBeginDocument{%
  \providecommand\BibTeX{{%
    \normalfont B\kern-0.5em{\scshape i\kern-0.25em b}\kern-0.8em\TeX}}}

\makeatletter
\preto{\@enddocumenthook}{}
\makeatother
\newenvironment{nospaceflalign*}
 {\setlength{\abovedisplayskip}{0pt}\setlength{\belowdisplayskip}{0pt}%
  \csname flalign*\endcsname}
 {\csname endflalign*\endcsname\ignorespacesafterend}

\begin{document}

\title[SciNoBo : A Hierarchical Multi-Label Classifier of Scientific Publications]{SciNoBo: A Hierarchical Multi-Label Classifier of Scientific Publications}

%%%%%%%%%%%%%%%%%%%%%%%%%%%%%%%%%%%%%%%%%%%%%%%%%%%%%%%%%%%%%%%%%
%%%%%%%%%%%%%%%%%%%%%%%%%%%%%%%%%%%%%%%%%%%%%%%%%%%%%%%%%%%%%%%%%
\author{Nikolaos Gialitsis}
\authornote{designed and developed the initial implementation and the theoretical formalism of \scinobo, undertook the literature review and the documentation of the proposed method.}
\orcid{0000-0002-3821-4851}
\affiliation{%
  \institution{Athena Research and Innovation Center}
  \department{Institute for Language and Speech Processing}
  \city{Athens}
  \country{Greece}
}
\email{ngialitsis@athenarc.gr}

%%%%%%%%%%%%%%%%%%%%%%%%%%%%%%%%%%%%%%%%%%%%%%%%%%%%%%%%%%%%%%%%%
%%%%%%%%%%%%%%%%%%%%%%%%%%%%%%%%%%%%%%%%%%%%%%%%%%%%%%%%%%%%%%%%%

\author{Sotiris Kotitsas}
\authornote{worked out almost all technical details, implemented the baseline method \dann, ran the experiments (including data collection and pre-processing) and comparative analysis,  documented and reviewed the results.}
\orcid{0000-0002-8114-6225}
\affiliation{%
  \institution{Athena Research and Innovation Center}
  \department{Institute for Language and Speech Processing}
  \city{Athens}
  \country{Greece}  
}
\email{sotiris.kotitsas@athenarc.gr}

%%%%%%%%%%%%%%%%%%%%%%%%%%%%%%%%%%%%%%%%%%%%%%%%%%%%%%%%%%%%%%%%%
%%%%%%%%%%%%%%%%%%%%%%%%%%%%%%%%%%%%%%%%%%%%%%%%%%%%%%%%%%%%%%%%%
\author{Haris Papageorgiou}
\authornote{conceived the study and supervised the project.}

\affiliation{%
  \institution{Athena Research and Innovation Center}
  \department{Institute for Language and Speech Processing}
  \city{Athens}
  \country{Greece}
}
\email{haris@athenarc.gr}
\date{}

%%%%%%%%%%%%%%%%%%%%%%%%%%%%%%%%%%%%%%%%%%%%%%%%%%%%%%%%%%%%%%%%%
%%%%%%%%%%%%%%%%%%%%%%%%%%%%%%%%%%%%%%%%%%%%%%%%%%%%%%%%%%%%%%%%%
\begin{abstract}
Classifying scientific publications according to Field-of-Science (FoS) taxonomies is of crucial importance, allowing funders, publishers, scholars, companies and other stakeholders to organize scientific literature more effectively. Most existing works address classification either at venue level or solely based on the textual content of a research publication. We present \scinobo, a novel classification system of publications to predefined FoS taxonomies, leveraging the structural properties of a publication and its citations and references organised in a multilayer network. In contrast to other works, our system supports assignments of publications to multiple fields by considering their multidisciplinarity potential. By unifying publications and venues under a common multilayer network structure made up of citing and publishing relationships, classifications at the venue-level can be augmented with publication-level classifications. We evaluate \scinobo on a publications' dataset extracted from Microsoft Academic Graph and we perform a comparative analysis against a state-of-the-art neural-network baseline. The results reveal that our proposed system is capable of producing high-quality classifications of publications. 
\end{abstract}

%%%%%%%%%%%%%%%%%%%%%%%%%%%%%%%%%%%%%%%%%%%%%%%%%%%%%%%%%%%%%%%%%
%%%%%%%%%%%%%%%%%%%%%%%%%%%%%%%%%%%%%%%%%%%%%%%%%%%%%%%%%%%%%%%%%


\maketitle

%%%%%%%%%%%%%%%%%%%%%%%%%%%%%%%%%%%%%%%%%%%%%%%%%%%%%%%%%%%%%%%%%
%%%%%%%%%%%%%%%%%%%%%%%%%%%%%%%%%%%%%%%%%%%%%%%%%%%%%%%%%%%%%%%%%
\section{Introduction} It is estimated that the overall volume of scientific publications doubles every 17.3 years, and every 12.9 years in physical and technical sciences~\cite{bornmann_growth_2021}.  In order to manage the wealth and growth rate of knowledge, multiple literature databases have emerged covering different perspectives: Microsoft Academic~\cite{wang_microsoft_2020}, Scopus~\cite{baas_scopus_2020}, Web Of Science~(WoS)~\cite{birkle_web_2020}, SemanticScholar, Crossref~\cite{hendricks_crossref_2020}, OpenCitations~\cite{peroni_opencitations_2020}, OpenAIRE~\cite{manghi_openaire_2019}, Dimensions~\cite{herzog_dimensions_2020}, ScienceDirect~\footnote{ScienceDirect [Internet]. Elsevier [1997] - [cited \today] 
Available from: https://www.sciencedirect.com/} as well as specialized databases such as PubMed~\footnote{PubMed [Internet]. Bethesda (MD): National Library of Medicine (US). [1946] - [cited \today]. Available from: https://www.ncbi.nlm.nih.gov/pubmed/} and the Computer Science Ontology (CSO)~\cite{salatino_computer_2018}.

%%%%%%%%%%%%%%%%%%%%%%%%%%%%%%%%%%%%%%%%%%%%%%%%%%%%%%%%%%%%%%%%%
%%%%%%%%%%%%%%%%%%%%%%%%%%%%%%%%%%%%%%%%%%%%%%%%%%%%%%%%%%%%%%%%%

To empower search engines (such as Google Scholar, Semantic Scholar, ArnetMiner~\cite{tang_arnetminer_2008}, InCites, Pub Finder~\cite{vergoulis_pub_2018}) recommendation systems~\cite{choochaiwattana_usage_2010}, science and innovation monitoring efforts~\cite{veugelers_impact_2015, grypari_research_2020, boaz_assessing_2009, stanciauskas_policy_2020, leal_filho_reinvigorating_2018}, and to normalize bibliometric indices~\cite{colledge_snowball_2014,colledge_scival_2014, ruiz-castillo_field-normalized_2015, vergoulis_bip_2021}, literature databases often adopt classification schemes for tagging entities (venues/publications/projects) in respect to their thematic contents: scientific areas, subject fields or topics. Entities can either be classified to one or more scientific fields from a predefined list,  or can be assigned a list of representative keywords by the author(s) or/and the journal editor(s). 

%%%%%%%%%%%%%%%%%%%%%%%%%%%%%%%%%%%%%%%%%%%%%%%%%%%%%%%%%%%%%%%%%
%%%%%%%%%%%%%%%%%%%%%%%%%%%%%%%%%%%%%%%%%%%%%%%%%%%%%%%%%%%%%%%%%

Scientific fields are often organized hierarchically into a taxonomy in which the top-levels represent broad subject areas and disciplines such as {\textit{social sciences}} and \textit{ engineering and technology} and the bottom levels represent fine-grained and specialized subfields such as \textit{optics \& laser technology}. Examples of taxonomies for scientific field classification are the: All Science Journal Classification (ASJC) System, International Classification of Diseases (ICD)~\cite{khoury_international_2017}, Frascati Manual Classification~\cite{oecd_frascati_2015}, Medical Subject Headings (MeSH)~\footnote{\href{https://meshb.nlm.nih.gov/treeView}{MeSH}}, WoS Categories and Subject Areas~\footnote{\href{https://images.webofknowledge.com/images/help/WOS/hp_research_areas_easca.html}{WoS}}, Scopus Subject Areas~\footnote{\href{https://service.elsevier.com/app/answers/detail/a_id/15181/supporthub/scopus/session/L2F2LzEvdGltZS8xNjM4Mzg3MzE1L2dlbi8xNjM4Mzg3MzE1L3NpZC9mVXg4b3ZPWldqbk9WRW9lUElKZWZGNk5YY0s4aGFBNno0cTNXelNVdDBUcDBwSDhMbXhMQ2dsaHlNa1dVbjFmaTIzbHRXellmMGFOMlUxajZHcDFIUnc2ZkJXNlNVWWNHOTJUNWFwcUNYS3BRQWo3dF96Q0kzZ0ElMjElMjE\%3D/}{Scopus Subject Areas}}, European Science Vocabulary (EuroSciVoc)~\footnote{\href{https://op.europa.eu/el/web/eu-vocabularies/euroscivoc}{EuroSciVoc}} and Microsoft Academic Graph Concepts~\cite{shen_web-scale_2018}. 

%%%%%%%%%%%%%%%%%%%%%%%%%%%%%%%%%%%%%%%%%%%%%%%%%%%%%%%%%%%%%%%%%
%%%%%%%%%%%%%%%%%%%%%%%%%%%%%%%%%%%%%%%%%%%%%%%%%%%%%%%%%%%%%%%%%

Recently, the scientometrics community has been shifting their focus from venue-level (journals/conferences) to publication-level classification systems, as evidenced by a growing line of research~\cite{shen_web-scale_2018, salatino_classifying_2018,waltman_new_2012, eykens_fine-grained_2021,  salatino_cso_2021, kandimalla_large_2021, rivest_article-level_2021, hoppe_deep_2021}. Comparative studies~\cite{perianes-rodriguez_comparison_2017,shu_comparing_2019} have shown that classification systems at the publication-level are more precise than venue-level counterparts as they naturally provide a higher degree of granularity, which can prove advantageous in certain applications. 
%%%%%%%%%%%%%%%%%%%%%%%%%%%%%%%%%%%%%%%%%%%%%%%%%%%%%%%%%%%%%%%%%
%%%%%%%%%%%%%%%%%%%%%%%%%%%%%%%%%%%%%%%%%%%%%%%%%%%%%%%%%%%%%%%%%

Despite the fact that publication-level systems have been proposed several decades ago~\cite{griffith_structure_1974, garfield_system_1975}, venue-level classifications, including WoS and Scopus~\cite{leydesdorff_global_2009, osborne_automatic_2016, archambault_towards_2011}, are still in use, since they are more easily curated by publishers and editors due to their smaller volume and more static nature. An extensive summary of approaches for venue-level classification and their limitations has been presented by~\citeauthor{archambault_towards_2011}~\cite{archambault_towards_2011}. 

%%%%%%%%%%%%%%%%%%%%%%%%%%%%%%%%%%%%%%%%%%%%%%%%%%%%%%%%%%%%%%%%%
%%%%%%%%%%%%%%%%%%%%%%%%%%%%%%%%%%%%%%%%%%%%%%%%%%%%%%%%%%%%%%%%%

The majority of methods proposed for automatic FoS categorization perform some form of clustering on publications in order to map science or to identify topics from scratch. These include, topic modeling approaches~\cite{he_detecting_2009,tang_arnetminer_2008} and the grouping of publications via citation networks~\cite{upham_emerging_2010, small_identifying_2014} or bibliographic coupling~\cite{waltman_new_2012}. Even though such unsupervised approaches are useful in many cases, they solve a different task to the one we focus on in this publication, which is the classifation of publications according to a predefined FoS taxonomy 

%%%%%%%%%%%%%%%%%%%%%%%%%%%%%%%%%%%%%%%%%%%%%%%%%%%%%%%%%%%%%%%%%
%%%%%%%%%%%%%%%%%%%%%%%%%%%%%%%%%%%%%%%%%%%%%%%%%%%%%%%%%%%%%%%%%

Our method \scinobo, contributes to the domain of taxonomy-driven FoS classification in more than one way. First and foremost, \scinobo classifies publications across all disciplines, in contrast to other works that focus on a specific domain. Secondly, it is suitable for handling multidisciplinarity as it can be applied in both multiclass and multi-label classification settings. Furthermore, \scinobo supports assignments to multiple levels of detail within a given FoS taxonomy by encoding hierarchical relationships among FoS labels. Moreover, \scinobo classifies publications by requiring minimal metadata. A publication can be classified from the first day it becomes available online, and later as more metadata gradually become available, \scinobo can classify the publication again by taking into account richer relationships. Lastly, we employ a new FoS taxonomy that extends OECD disciplines with \sciencemetrix FoS codes.
%%%%%%%%%%%%%%%%%%%%%%%%%%%%%%%%%%%%%%%%%%%%%%%%%%%%%%%%%%%%%%%%%
%%%%%%%%%%%%%%%%%%%%%%%%%%%%%%%%%%%%%%%%%%%%%%%%%%%%%%%%%%%%%%%%%

The publication is structured as follows: We start by reviewing different approaches found in the literature for classifying publications according to a predefined FoS taxonomy. In section 3, we discuss the main limitations of existing approaches and the drivers for additional research. We then proceed in section 4 and describe in-detail the proposed methodology and its mathematical formulation. In the next section, we report on conducted experiments and implementation details. Next, we present results and discuss how \scinobo performs against a neural state-of-the-art baseline. Finally, we reach conclusions and propose directions for subsequent research and improvements.
%%%%%%%%%%%%%%%%%%%%%%%%%%%%%%%%%%%%%%%%%%%%%%%%%%%%%%%%%%%%%%%%%
%%%%%%%%%%%%%%%%%%%%%%%%%%%%%%%%%%%%%%%%%%%%%%%%%%%%%%%%%%%%%%%%%

\section{Related Work}
\label{relatedwork}
Approaches for publication-level FoS classification mostly rely on metadata including titles, author-keywords and abstracts, since full text is often unavailable or locked behind a paywall. We distinguish two main approaches found in literature: keyword extraction methods, and machine learning methods. 
%%%%%%%%%%%%%%%%%%%%%%%%%%%%%%%%%%%%%%%%%%%%%%%%%%%%%%%%%%%%%%%%%
%%%%%%%%%%%%%%%%%%%%%%%%%%%%%%%%%%%%%%%%%%%%%%%%%%%%%%%%%%%%%%%%%

\subsection{Keyword extraction methods}{

Keyword extraction methods have been applied with the goal of identifying small sets of representative words, phrases, or n-grams to associate with a predefined set of FoS. According to the similarity scores (e.g. measuring overlap, or some vector space similarity such as the angle between word-vectors), the FoS candidates are ranked and the publication is classified to the best matching field(s). 

For example,~\citeauthor{salatino_cso_2021}~\cite{salatino_cso_2021} proposed a text-based classifier for classifying publications to one or more research area(s) from the Computer Science Ontology (CSO~\cite{salatino_computer_2018}). N-grams extracted from each abstract are matched to FoS labels by means of Levenshtein similarity as well as by the cosine similarity between their pre-trained word2vec embeddings. Even though their approach performs multi-label classification and supports hierarchical assignments through CSO child-parent relationships, intensive post-processing effort is needed to detect and filter-out false-positives. Also, the classifications are confined to the Computer Science domain.

\citeauthor{shen_web-scale_2018}~\cite{shen_web-scale_2018} describe the concept tagging of publications within Microsoft Academic Graph (MAG). Publications are represented based on both graph structural and textual information by leveraging metadata including venue names, titles, keywords and abstracts, in addition to the metadata of their neighbors within the graph. Similarly, MAG concepts are represented by the first paragraph of their corresponding Wikipedia entry (concepts are derived from Wikipedia). The vector embeddings of both representations are compared through cosine similarity and the publication is classified to the concept (FoS) if the similarity exceeds a predefined threshold. Nevertheless, the representation of a MAG concept directly depends on its Wikipedia entry, whereas for most taxonomies, only word labels represent the classes. Furthermore, empirical weights and heuristics are applied therein which prohibits complete reusability.

}
\subsection{Machine learning methods}{
 \label{ml}

\subsubsection{Traditional methods}
\label{traditional-ml}
This category encompasses supervised machine learning methods i.e. labeled examples are required in order to train the model. These were some of the first approaches towards the automatic classification of publications according to a pre-defined FoS taxonomy.

\citeauthor{caragea_co-training_2015}~\cite{caragea_co-training_2015} classify publications to one out of six FoS categories from the CiteSeer literature database by taking advantage of citation relationships. The citation contexts of publications (citing and cited) are used in order to train different variations of the Multinomial Naive Bayes classifier. However, this approach does not support multi-label assignments and is only tested on a classification scheme containing few non-hierarchical FoS categories. 

Domain-specific FoS classification systems have also been sporadically developed. {\citeauthor{lukasik_hierarchical_2013}~\cite{lukasik_hierarchical_2013}} examine a combination of Naive Bayes and kNN algorithm on the hierarchical multi-label classification of publications according to Mathematical Subject Headings (MSC). Similarly,~\citeauthor{kurach_multi-label_2013}\cite{kurach_multi-label_2013} construct classifier ensembles in order to assign MeSH terms to biomedical publications from the Pubmed Central database. They evaluate various combinations of machine learning methods in a supervised multi-label classification setting. However, their study does not account for hierarchical relationships among FoS labels.

 \subsubsection{Deep learning methods}
 \label{sssec:deeplearning}
Nowadays, traditional machine learning approaches for FoS classification have mostly been superseded by deep learning methods.
 
 \citeauthor{kandimalla_large_2021}~\cite{kandimalla_large_2021} employ a deep neural encoder equipped with the attention mechanism in order to classify publications to WoS subject categories. They keep the most representative words per abstract in terms of \tfidf score and subsequently map them to pre-trained word embeddings before feeding them to the neural network. However their approach does not take into account the hierarchical relationships of the FoS taxonomy and finds difficulty in discriminating between FoS categories with similar vocabulary.

In~\cite{rivest_article-level_2021}, a modified character-level Convolutional Neural Network (CNN) is compared against the more traditional approaches of bibliographic coupling and direct citation for the problem of FoS classification. They evaluated the performance of each method separately on a publication-level human-curated test dataset comprising of 200 publications in total, half randomly sampled from disciplinary journals, and the other half from multidisciplinary. For training the CNN, the training labels used were direct extensions of the venue label. No clear winning method was announced in the comparative study as all resulted in similar performance. Multiple sources of metadata were given as input to the CNN, including authors' affiliations, referenced venue names, reference titles, abstract, keywords, title, and subfield classification of references. Most of these however are rarely available. Furthermore, the text-based input is concatenated/truncated arbitrarily and does not take into account term semantics. 

 \citeauthor{hoppe_deep_2021}\cite{hoppe_deep_2021} combine a BiLSTM neural network architecture with Knowledge Graphs (KG) in order to represent class labels by their KG-embeddings. The intuition is that by including the KG embeddings of the classes, in addition to their word labels, the FoS classifier will achieve higher performance. They demonstrate the above, on the multi-label classification of $\sim92K$ ArXiv publications represented by DBpedia embeddings. Nevertheless, the relationship between the FoS labels is not taken into acount as the multi-label classification problem is decomposed into independent binary classifications.
 
 In the position paper HierClasSArt~\cite{alam_hierclassart_2021}, knowledge graphs (KGs) were proposed, in addition to neural networks, for the FoS classification of publications to mathematical topics from the zbMATH database. KG-embeddings leveraging both  textual and structural metadata can be derived from the available metadata, and successively be provided as input to a deep neural network. Hierarchical relations among entities may be captured through the help of NLP methods or human experts, but the exact details are not thoroughly described. While such a methodology can potentially be transferred to other domains, only the domain of mathematics was explored in the study.

}
\section{Motivation}
\label{motivation}
Existing FoS classification approaches completely ignore, or face significant difficulties when dealing with multidisciplinarity, either at the venue or at the publication-level. Moreover, nearly all of them depend on textual content which, when available, is prone to concept drift, discourse norms in specific fields and multilinguality specificities. Moreover, several approaches confine classifications to a specific discipline or lack generalization capabilities. In addition, hierarchical relationships between FoS labels are often under-utilized or ignored.

Therefore, there is still a need for developing systems for efficient multi-label classification at the publication level. Our motivation is that by taking into account both the citing/publishing relationships at the publication-level as well as at the venue-level, we will be able to provide "context-aware" classifications without considering  publication content. In contrast to other methods, we propose a multiclass multi-label classification approach assigning research publications to one or more FoS codes capturing the increasing multidisciplinarity in literature.
\begin{figure}[!t]
\centering
\includegraphics[width=0.5\textwidth]{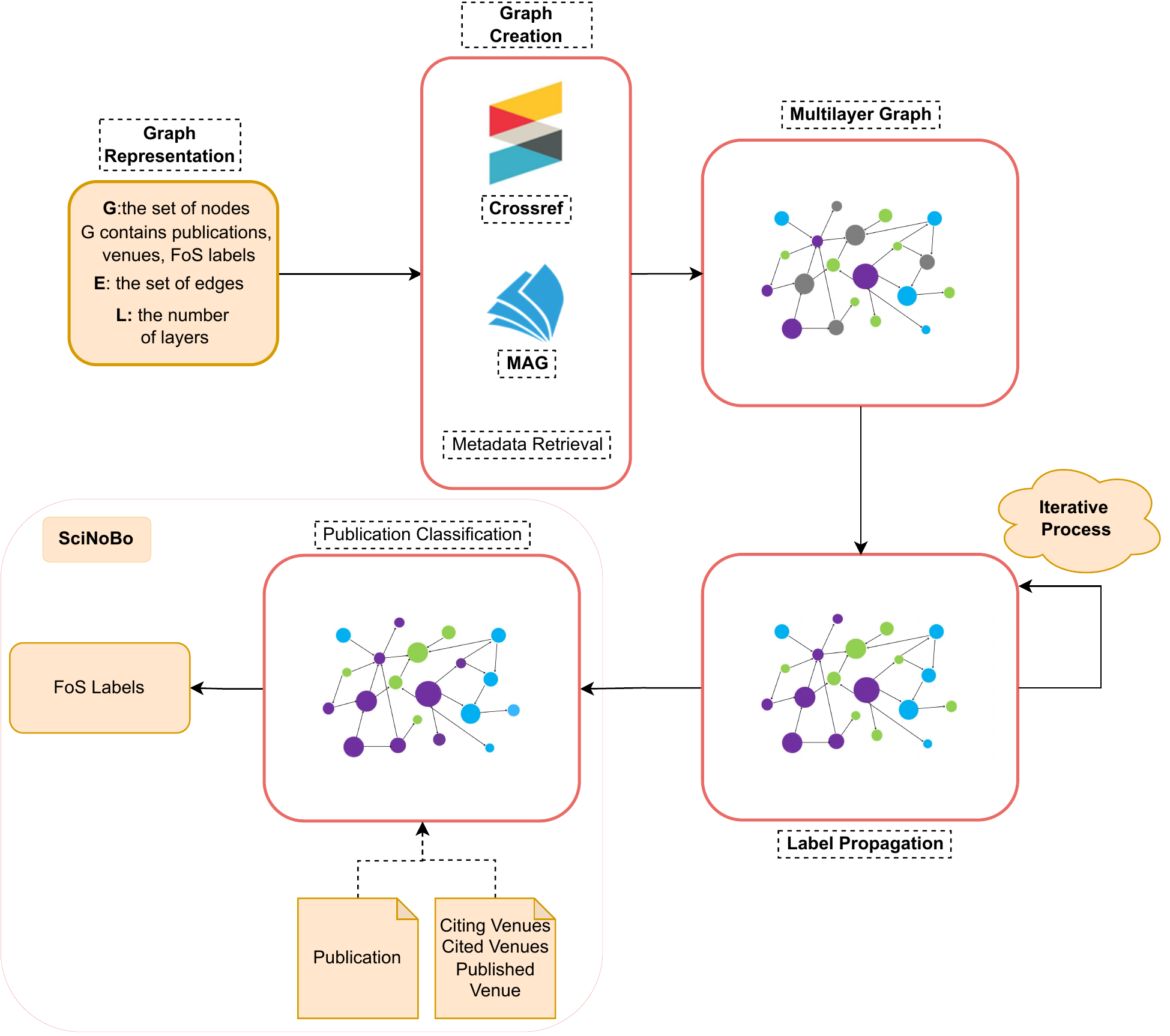}
\caption{Illustration of the proposed method. After we define our graph in Graph Representation, we retrieve metadata and construct it. The result is a Multilayer Graph and after Label Propagation, we can input a publication along with the required metadata to retrieve the FoS labels.}
% \vspace*{-4mm}
\Description[the methodology decomposed into separate steps]{The steps of the methodology consist of a graph representation step followed by a graph creation step resulting in a multilayer graph, an iterative label propagation within the graph, and finally a publication classification step which takes as input the graph, the publication and metadata information, and outputs FoS labels}
\label{scinobo_pipeline}
\end{figure}

\section{SciNoBo}
\label{methodology}
The method we propose is based on the assumption that a publication~\footnote{We use the term "publication" to refer to all peer-review research works published in journals or conferences.} mostly cites thematically related publications. We bridge venues (journals/conferences) and publications by constructing a multilayer network (graph) in which venues are represented by nodes, and venue-venue edges reflect citing-cited relationships in their respective publications. \scinobo classifies publications to one or more FoS based on the publishing venues of the publications it references (out-citations) and the publishing venues of the publications it gets cited by (in-citations). Therefore, \scinobo classifies publications with minimal metadata utilizing only journal or conference names as well as citing information. Figure~\ref{scinobo_pipeline} illustrates the steps followed to create \scinobo. Each step of the approach is analytically covered in the following subsections.

\subsection{Graph Representation}
\label{representation}

\begin{figure}[t]
\centering
\includegraphics[width=0.3\textwidth]{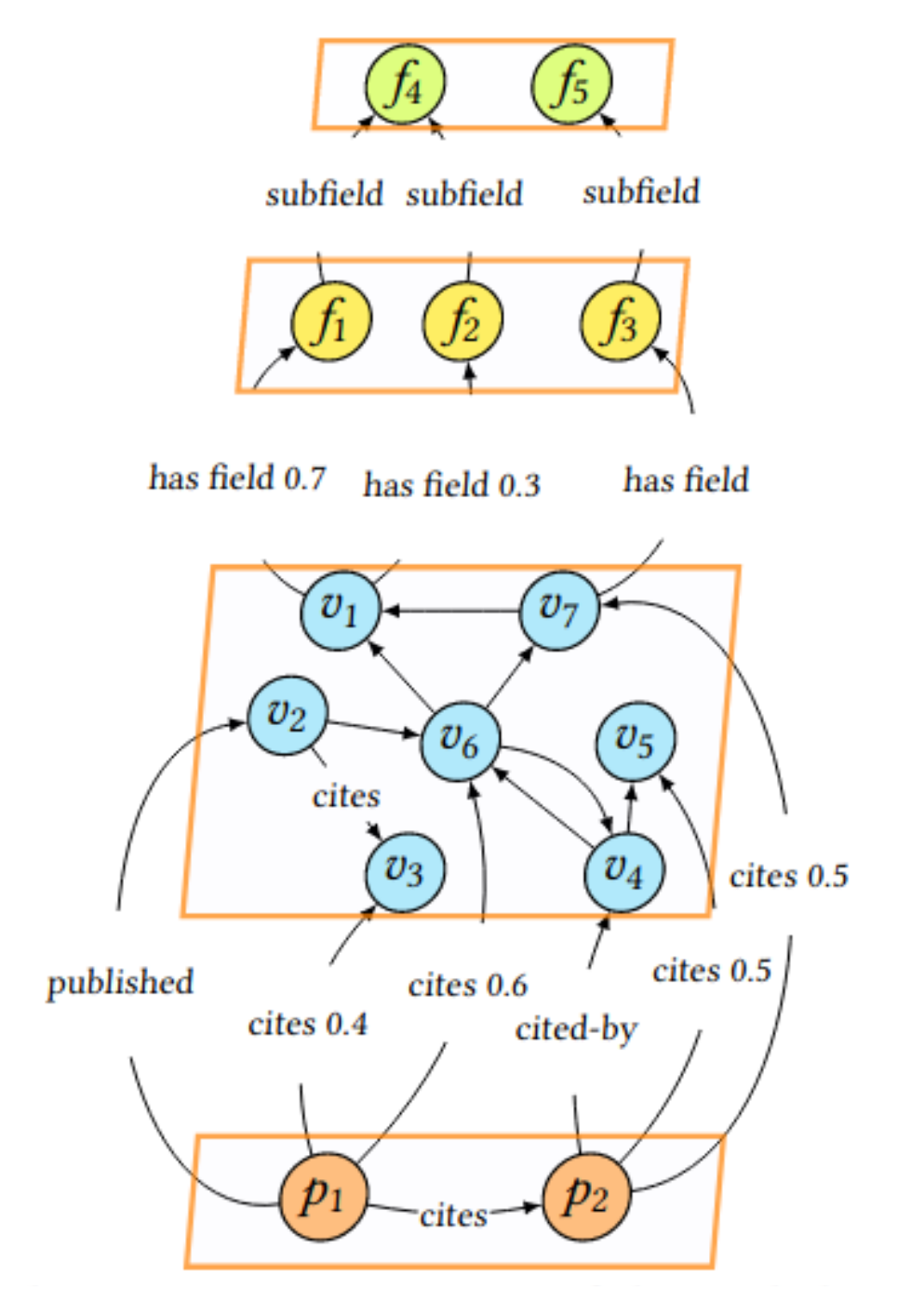}

\Description[Multiple layers of the network and the connections between them]{Multiple layeers of the network: the first layer contains publications, the second layer contains venues, the third layer contains fields of science, and the forth layer contains more broad fields of science. Each layer is connected to the next one through edges between its nodes}
\caption{Schematic representation of the multilayer network}
\label{fig:scinobo-schema}
\end{figure}

\scinobo unifies multiple types of relationships (edges) between entities as well as multiple types of entities under a common framework of operations represented as a multilayer network~\footnote{Multilayer networks are data
structures used to model complex interactions~\cite{kivela_multilayer_2014},
ranging from Biomedicine~\cite{hammoud_multilayer_2020}, to Social Network Analysis ~\cite{contractor_network_2011}.}~(see Figure \ref{fig:scinobo-schema}). We consider a multilayer network $\mathcal{G} = (\mathcal{V}, \mathcal{E}, \mathcal{L})$ where $\mathcal{V}$ contains the set of publications $\mathcal{P}$, the set of venues $\mathcal{J}$, and the set of scientific fields $\mathcal{F}$. Or equivalently: $\mathcal{V} = \mathcal{P}\bigcup\mathcal{J}\bigcup\mathcal{F}$. The symbol $\mathcal{E}$ represents the set of edges (links) between nodes; and $\mathcal{L}$ is the set of layers capturing different types of relationships between nodes. Since the network has multiple layers, each edge belongs to one layer $\ell \in \mathcal{L}$ and has a weight $w \in \mathbb{R}^+$. We can represent all edges in the network using 4-tuples as: $\mathcal{E} = \{(u, v, \ell, w); u,v \in \mathcal{V}, \ell \in \mathcal{L}, w \in \mathbb{R}^+\}$ and $(u, v, \ell, w) \neq (v, u, \ell, w)$ (directed edges). Edges in layer $\ell \in \mathcal{L}$ represent a particular type of connection among nodes, and two nodes $u, v$ might be connected by edges in multiple layers.

We formulate the task of scientific field classification as a link-prediction problem in the multilayer network. The goal is to predict edges between publication-nodes and scientific-field nodes: $\{(u, v, \ell_0, w); u \in \mathcal{P}, v \in \mathcal{F}, \ell_0 \in \mathcal{L}, w \in \mathbb{R}^+\}$.

An edge between two publications $(p_i,p_j)$ at $\ell_1$ means that publication $p_i$ cites publication $p_j$.  An edge at layer $\ell_2$ connects a publication to its publishing venue(s). We also define edges between venues at $\ell_3$: $\{(u, v, \ell, w); u,v \in \mathcal{J}, \ell \in \mathcal{L}, w \in \mathbb{R}^+\}$; where $w$ is the number of publications which have been published in venue $u$ and cite (reference) publications published in venue $v$. The weight of an edge $(v,f)$ at $\ell_4$ corresponds to how thematically relevant the publications published in $v$ are to the scientific field $f$. Finally, $\ell_5+$ layers represent hierarchical relationships among labels.

\subsection{Graph Creation}
\label{graph_creation}
\scinobo network was populated by exploiting \crossref~\footnote{https://www.crossref.org/} and Microsoft Academic Graph~(\acmag)~\footnote{ https://www.microsoft.com/en-us/research/project/microsoft-academic-graph/
}. \crossref contains more than 120 million publications and \acmag contains approximetaly 250 million records. We retrieve all the publications that were published between $2016-2021$, along with their references ~\footnote{references retrieved were confined in a 10 year window. Older references should be omitted as it is customary for research papers to cite old works (e.g. in the introduction section) in order to provide general background knowledge whose themes may not directly coincide with the publication's.} and their citations when available. 

For every publication, the publishing venue is contained in the metadata. However this is not the case for the references and citations. As a result, for every publication we query its references and citations in \crossref/\acmag (by taking the union of the metadata) and we retrieve the original metadata of the reference or the citation. Inherently, we can now create venue-venue relationships (edges at layer $\ell_3$) as in~\ref{representation}. The weight of a venue-venue edge is the amount of times a venue has referenced or being cited by another venue. An edge is created between two venues if they (i.e. their respective publications) cite each other more than 10 times~\footnote{We observed that by using a lower weight threshold we introduce noise into the network by including low-signal venues of questionable quality. On the contrary, a weight threshold $t \sim 100$ leads into a large number of reputable conferences to being omitted as their individual instances (e.g. annual) are filtered before the venue deduplication and aggregation steps, discussed in (\ref{deduplication}), take place.}. Post graph creation, we normalize the weights of each node's outgoing edges to sum up to $1$ by diving with the maximum weight of each neighborhood. The normalized weight of a venue-venue edge ($u,v$) can be roughly interpreted as the probability of a publication published in $u$ to cite a publication published in $v$.

\paragraph{\textbf{Venue Deduplication}}\label{deduplication}: A considerable challenge is dealing with naming inconsistencies in the reporting of venues in publication references/citations, or different instances of the same venue. This challenge is particularly prevalent in Crossref metadata since the published venue of each publication is being deposited by the authors. Our main goal is to create abbreviations for the names of the venues e.g. the "Empirical Methods in Natural Language Processing" conference should be mapped to \emnlp. Furthermore, different instances of venues should also be mapped to a unique venue abbreviation (e.g. \emnlp $2019$, \emnlp $2020$ etc. to \emnlp)~\footnote{The following preprocessing was applied to the names of the venues: Removal of latin, cardinal and ordinal numbers, dates, days, months, pre-specified  words/phrases (e.g. "speech given", "oral talk" etc.), stopwords, special characters;adding a space when removing them and normalising multiple spaces. The same preprocessing procedure is also applied during inference.}. In addition by performing an exploratory analysis on the names of the reported venues, we conclude that most of the abbreviations exist after the character '$-$' and inside parentheses.

While parsing the publications during the Graph Creation process, we keep a mapping from the full venue names to the venue abbreviations we have identified, while creating the venue-venue edges. The mapping created is also used during inference time, to map the incoming venues to the abbreviated venue names in the multilayer graph of \scinobo.

\paragraph{\textbf{Field-of-Science Taxonomy}}:
\label{fos_ontology}
Our classification scheme is underpinned by the OECD disciplines/fields of research and development (FORD) classification scheme, developed in the framework of the Frascati Manual~\footnote{https://www.oecd.org/sti/inno/frascati-manual.htm} and used to classify R{\&}D units and resources in broad (first level(L1), one-digit) and narrower (second level(L2), two-digit) knowledge domains based primarily on the R{\&}D subject matter. To facilitate a more fine grained analysis, we extend the OECD/FORD scheme by manually linking FoS labels of the \sciencemetrix~\footnote{https://science-metrix.com/} classification scheme to OECD/FORD level-2 categories, creating a hierarchical 3-layer taxonomy. Table~\ref{tab:fos_statistics} provides stastistics of our taxonomy.
\begin{table}[h]
\begin{tabular}{@{}ll@{}}
\toprule
\textbf{Levels of FoS} & \textbf{Number of Labels} \\ \midrule
Level 1                & 6                             \\
Level 2                & 42                            \\
Level 3                & 174                           \\ \bottomrule
\end{tabular}
% \vspace*{3mm}
\caption{Statistics of the extended OECD/FORD classification scheme.}
\label{tab:fos_statistics}
\end{table}

\sciencemetrix Classification also provides a list of Journal Classifications. We integrate this list, by mapping its journals to \scinobo nodes and linking them with the relevant FoS codes. This mapping represents $\ell_4$ relationships, which are utilized to classify publications in FoS labels. Initially a small portion of venues have an FoS in Level-2 and Level-3. By utilizing Label Propagation, we aim to increase the venue label coverage.

\subsection{Label Propagation}
\label{propagation}

The intuition behind incorporating venues into the network, is that  starting from a small set of seeds (venues with FoS labels), we can propagate the information to the rest of the network. The hypothesis is that a venue is more likely to express the FoS of its most referenced venues, an approach resembling a nearest-neighbor classification setting. 

We assume that only a subset of venues $\mathcal{J}^*\subseteq\mathcal{J}$ has available FoS labels (i.e. venue-FoS edges in layer $\ell_4$). However, we do not consider these seed venue-FoS classifications to be ground-truth and we re-evaluate them dynamically during label propagation. By taking into account the network of venue-venue relationships, we enrich the initial FoS journal classifications described in \ref{fos_ontology} by inferring additional venue-FoS relationships. Consequently, previously single-labeled classifications may become multi-labeled after a few rounds of label propagation. 

Label propagation is an iterative procedure. On each iteration, every venue-node aggregates the FoS labels of its neighbors in proportion to the venue-neighbor preference and neighbor-FoS preference. The preference of a venue towards one of its neighbors is expressed through the venue-venue normalized edge weight at $\ell_4$. A weight of $0$ equals to no preference (i.e. publications do not cite publications published by the neighbor venue) whereas a weight of $1$ equals to absolute preference (all publications cite publications published by the neighboring venue). These two preferences can be multiplied together, to estimate the expected weight between the venue, and each FoS node connected to its neighbors. 

\text{Given layers:}$\ell_t, \ell_{t+1} \text{the label propagation formula is:}$
\begin{nospaceflalign*}
\stackrel{\mathclap{\mbox{venue-FoS}}}{w_{i, k}^{\ell_{t+1}}}  =  \sum
\limits_{j \in \mathscr{N}_{i}^{\ell_t}
}^{}  \sum
\limits_{k \in \mathscr{N}_{j}^{\ell_{t+1}}
}^{} 
\stackrel{\mathclap{\mbox{venue-neighbor}}}{w_{i, j}^{\ell_t}}\qquad  \cdot \qquad   \stackrel{\mathclap{\it{\mbox{neighbor-FoS}}}}{w_{j, k}^{\ell_{t+1}}}\\
\stackrel{\mathclap{\mbox{for all citing venues in $\ell_t$}}}{\forall{i \in \{ u \colon\  \exists(u, \cdot, \ell_t, \cdot); u \in \mathcal{V}\}}}\\
\stackrel{\mathclap{\mbox{for each cited neighbor's FoS}}}{\forall{k \in \{ v \colon\  \exists(j, v, \ell_{t+1}, \cdot); j \in \mathscr{N}_{i}^{\ell_t} ; {v \in \mathcal{V}} \}}} \\
\text{where }\mathscr{N}_{i}^{\ell_t} \text{ are venues cited by } i \in \ell_t
\end{nospaceflalign*}

Complete statistics regarding the Graph (before and after Label Propagation) are presented in Table~\ref{tab:graph_statistics}. We observe that Label Propagation achieves wide coverage between venue-FoS ($\ell_4$) edges.
                
\begin{table}[!t]\small
    \begin{tabular}{@{}cccccccc@{}}
        \toprule
        Nodes &
        Edges &
          \multicolumn{3}{c}{\begin{tabular}[c]{@{}c@{}}Pre Label Propagation\\ Lvl1-Lvl2-Lvl3\end{tabular}} &
          \multicolumn{3}{c}{\begin{tabular}[c]{@{}c@{}}Post Label Propagation\\ Lvl1-Lvl2-Lvl3\end{tabular}} \\ \midrule
          94752 &
          3112953 &
          N/A &
          84 &
          14889 &
          32049 &
          32049 &
          32324 \\ \bottomrule
    \end{tabular}
\caption{\small{Complete graph statistics. Pre Label Propagation indicates the number of venues that have FoS labels per Level before Label Propagation. Post Label Propagation indicates the number of venues with FoS Labels per Level after we apply Label Propagation. Nodes represent the number of venue nodes in the multilayer graph.}}
% \vspace{-1mm}
\label{tab:graph_statistics}
\end{table}

\subsection{Publication Classification}
\label{inference}
\scinobo can assign FoS labels to individual publications through \it{citing/cited-by relationships} and \it{publishing/published-by relationships}. \normalfont Publication-classification uses the same label propagation mechanism as the one presented in \ref{propagation}. Assume that each publication is represented by a unique node in the \scinobo network. The goal is to connect each publication node to one or more FoS nodes in layer $\ell_0$. We have already discussed how venue-FoS relationships ($\ell_4$) can be established in the subsections \ref{fos_ontology} and \ref{propagation}. 

There exist multiple ways to back-propagate information from the venue level to the publication level depending on the available metadata:

\begin{enumerate}[noitemsep,topsep=-1pt]
    \item based on the published venues ({\it{Published-by}})
    \item based on the  referenced/cited venues ({\it{References}}) 
    \item based on the cited+citing venues ({\it{References + Citations}})
\end{enumerate}

% \smallskip
\paragraph{\textbf{Published-by}}(\scinobopub):
\label{published-by}
Given a publication $p$ and the set of distinct venues in which it has been published in $\{v_1,v_2,\cdots,v_n\}$ we draw weighted edges in layer $\ell_2$ of weight $w_{p,v}= \frac{1}{n}$. Thus, the weight is equally distributed among all published venues. Consequently, each venue contibutes a score $\frac{w_{venue, FoS}}{n}$ to the publication. The scores per FoS are aggregated and ranked according to their total weights. The publication is finally classified to the top $T$ FoS, where $T$ might be fixed or be equal to the number of weights that exceed a user-defined threshold.

\paragraph{\textbf{References}}(\scinoboref):
\label{references_approach}
Given a publication $p$ and a set of venues it references $\{v_1,v_2,\cdots,v_n\}$ we draw edges in layer $\ell_2$ of weight $w_{p,v}=\text{(number of referenced publications published in $v$)/$n$}$. Each venue contributes a score $(w_{p,v}) \cdot (w_{venue,FoS})\ $ where $w_{venue,FoS}$ is the normalized weight of the venue-FoS edge in $\ell4$. Similar to the ({\it{published-by approach}}), the weights are aggregated and the publications are assigned to the top $T$ FoS.

\paragraph{\textbf{References + Citations}}(\scinobocitref): 
This approach is identical to the reference approach except that in addition to referenced venues, the cited-by (citation) venues are also included in the venue set if available. A methodology originally proposed in the context of one particular field might eventually prove ground-breaking in a completely different field. By incorporating citation venues, \scinobo captures cross-domain FoS that would otherwise be missed.
%%%%%%%%%%%%%%%%%%%%%%%%%%%%%%%%%%%%%%%%%%%%%%%%%%%%%%%%%%%%%%%%%%%
\section{Experiments}
\label{experiments}
%%%%%%%%%%%%%%%%%%%%%%%%%%%%%%%%%%%%%%%%%%%%%%%%%%%%%%%%%%%%%%%%%%%
%%%%%%%%%%%%%%%%%%%%%%%%%%%%%%%%%%%%%%%%%%%%%%%%%%%%%%%%%%%%%%%%%%%

\subsection{Dataset}
\label{dataset}
% For our main experiments, where we predict Field-of-Science (FoS) categories,
In our experiments, we utilize the \sciencemetrix Journal classification. \sciencemetrix provides a list of Journals alongside with FoS labels~(\ref{graph_creation}). Furthermore, these FoS categories have been mapped to Level-3 FoS categories in our taxonomy~(\ref{fos_ontology}).
%%%%%%%%%%%%%%%%%%%%%%%%%%%%%%%%%%%%%%%%%%%%%%%%%%%%%%%%%%%%%%%%%%%
%%%%%%%%%%%%%%%%%%%%%%%%%%%%%%%%%%%%%%%%%%%%%%%%%%%%%%%%%%%%%%%%%%%

\begin{figure}[t]
\centering
\includegraphics[width=0.5\textwidth]{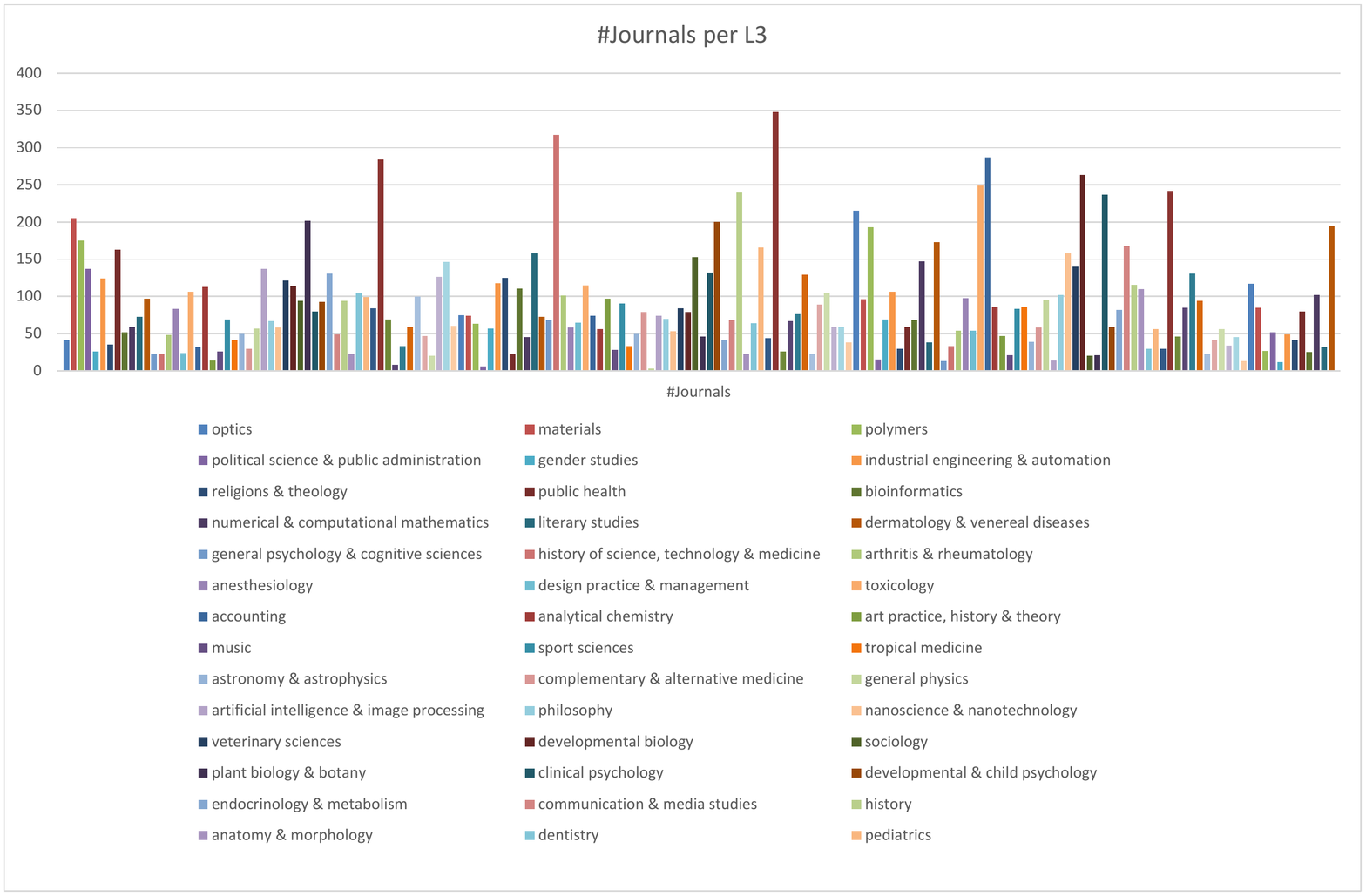}
\Description[histogram of journals per L3 Field of Science]{A  histogram of journals per L3 field of science category}
\caption{Distribution of Journals, classified from \sciencemetrix into Level-3 FoS categories in our taxonomy. Names of all Level 3 FoS labels were ommited for simplicity.}
% \vspace*{-4mm} 
\label{journal_distribution}
\end{figure}

To create, a comprehensive, large-scale, and clean dataset, we retrieve publications from Microsoft Academic Graph (\acmag) that are published in the Journals classified from \sciencemetrix. \acmag provides a wide range of publications. Figure~\ref{journal_distribution} presents the number of Journals that \sciencemetrix has classified to Level-3 FoS in our taxonomy. One can easily observe, that by retrieving a certain amount of publications per journal, an unbalanced dataset will be created. We retrieve 500 publications per Journal and per Level 3 FoS. The unbalanced dataset created, describes real-world data, hence our evaluation splits follow this unbalanced distribution.
%%%%%%%%%%%%%%%%%%%%%%%%%%%%%%%%%%%%%%%%%%%%%%%%%%%%%%%%%%%%%%%%%%%
%%%%%%%%%%%%%%%%%%%%%%%%%%%%%%%%%%%%%%%%%%%%%%%%%%%%%%%%%%%%%%%%%%%

Moreover, we compare \scinobo to a deep learning method, which requires a balanced train dataset. To that end, we further sample \acmag to obtain 10K train samples per Level 3 FoS. The final dataset statistics are presented in Table~\ref{dataset_statistics}. For every publication, we also retrieve abstracts and additional metadata. 
%%%%%%%%%%%%%%%%%%%%%%%%%%%%%%%%%%%%%%%%%%%%%%%%%%%%%%%%%%%%%%%%%%%
%%%%%%%%%%%%%%%%%%%%%%%%%%%%%%%%%%%%%%%%%%%%%%%%%%%%%%%%%%%%%%%%%%%

One limitation of the above-mentioned approach, is that the \sciencemetrix classification is at the journal-level and not at the publication-level. We try to mitigate this effect by excluding multidisciplinary (e.g \plosone) journals and assume that the journal-level classification also represents the publication-level classification.
\begin{table}[!h]\small
\begin{tabular}{@{}lllll@{}}
\toprule
Statistics          & Train Set & Development Set & Test Set & Total   \\ \midrule
Number of Instances & 1687826 & 120282          & 120307   & 1928415 \\ \bottomrule
\end{tabular}
\caption{\small{Statistics of the dataset used for training and evaluating our methods.}}
\label{dataset_statistics}
\end{table}

%%%%%%%%%%%%%%%%%%%%%%%%%%%%%%%%%%%%%%%%%%%%%%%%%%%%%%%%%%%%%%%%%%%
%%%%%%%%%%%%%%%%%%%%%%%%%%%%%%%%%%%%%%%%%%%%%%%%%%%%%%%%%%%%%%%%%%%

\subsection{Baseline Method}
\label{baseline_method}
We compare \scinobo, to \dann~\cite{kandimalla_large_2021}, a state-of-the-art model which utilizes textual information (abstracts) with a deep attentive neural network. \dann~\cite{kandimalla_large_2021}, represents each abstract, by sorting the words based on \tfidf scores, keeping the top $d$ words, reverting to the original ordering and encoding them with pretrained Word Embeddings. Then, a deep neural encoder, with a bidirectional \rnn and \selfattention, encodes the abstract and a final softmax layer outputs the probabilities for each FoS category.

%%%%%%%%%%%%%%%%%%%%%%%%%%%%%%%%%%%%%%%%%%%%%%%%%%%%%%%%%%%%%%%%%%%
%%%%%%%%%%%%%%%%%%%%%%%%%%%%%%%%%%%%%%%%%%%%%%%%%%%%%%%%%%%%%%%%%%%

Despite being one of the first works that utilized Deep Neural Networks, along with textual information to classify publications into FoS categories, the \tfidf filtering of the words of the abstracts breaks the sequence order and limits the effects of the \rnn encoders. Furthermore, \dann cannot perform hierarchical classification, whereas our approach is inherently hierarchical, since the \scinobo network can provide publication-FoS codes at all 3 Levels of the taxonomy by employing the label propagation mechanism described in ~\ref{propagation} and ~\ref{inference}.
%%%%%%%%%%%%%%%%%%%%%%%%%%%%%%%%%%%%%%%%%%%%%%%%%%%%%%%%%%%%%%%%%%%
%%%%%%%%%%%%%%%%%%%%%%%%%%%%%%%%%%%%%%%%%%%%%%%%%%%%%%%%%%%%%%%%%%%

\subsection{Implementation Details}
\label{implementation_details}
Following~\cite{kandimalla_large_2021}, we employ their best model, which utilizes pretrained \fasttext embeddings on a corpus created from WoS and \tfidf filtering of the abstracts by keeping the $d=80$ most relevant words. To that end, we created a large corpus by retrieving publications from \acmag, assigned to Level 3 FoS labels in our taxonomy. The created corpus contains approximetaly 20 million publications with abstracts. We trained \fasttext embeddings and calculated \tfidf scores. Regarding, the neural encoder of \dann, we employed our pretrained \fasttext embeddings with dimensionality of 100 and utilized 128 neurons at the recurrent encoders. Dropout was set at $0.2$ and \adam was used as the optimizer with learning rate of $10^{-3}$. We utilize Early Stopping with patience of 20, perform multiple experiments to account for standard deviation and present averaged results.
%%%%%%%%%%%%%%%%%%%%%%%%%%%%%%%%%%%%%%%%%%%%%%%%%%%%%%%%%%%%%%%%%%%
%%%%%%%%%%%%%%%%%%%%%%%%%%%%%%%%%%%%%%%%%%%%%%%%%%%%%%%%%%%%%%%%%%%

Regarding \scinobo, no hyperparameter selection is required. After the Label Propagation (\ref{propagation}) procedure is finished, the only information needed to infer publications to FoS categories, are published and citing/cited venues as stated in~\ref{inference}. We perform 2 rounds of Label Propagation.
%%%%%%%%%%%%%%%%%%%%%%%%%%%%%%%%%%%%%%%%%%%%%%%%%%%%%%%%%%%%%%%%%%%
%%%%%%%%%%%%%%%%%%%%%%%%%%%%%%%%%%%%%%%%%%%%%%%%%%%%%%%%%%%%%%%%%%%

\subsection{FoS Classification \& Evaluation}
\label{fos_classification_evaluation}
Given that \dann~(\ref{baseline_method}) cannot perform hierarchical classification, evaluation was carried out at level-L3 of our classification scheme (i.e., 174 FoS Labels).
%%%%%%%%%%%%%%%%%%%%%%%%%%%%%%%%%%%%%%%%%%%%%%%%%%%%%%%%%%%%%%%%%%%
%%%%%%%%%%%%%%%%%%%%%%%%%%%%%%%%%%%%%%%%%%%%%%%%%%%%%%%%%%%%%%%%%%%
In Section~\ref{inference}, the different approaches of classifying publications with \scinobo have been explored. To analyze the impact of each classification approach, we present results for each variant. Evidently, this analysis can also be viewed as an ablation study. 
%%%%%%%%%%%%%%%%%%%%%%%%%%%%%%%%%%%%%%%%%%%%%%%%%%%%%%%%%%%%%%%%%%%
%%%%%%%%%%%%%%%%%%%%%%%%%%%%%%%%%%%%%%%%%%%%%%%%%%%%%%%%%%%%%%%%%%%

The evaluation dataset and the baseline (\dann) support only multiclass classification, whereas \scinobo can be utilized in multilabel and multiclass tasks. Publication-level classification cannot always be addressed as a multiclass task, since a growing number of multidisciplinary publications is published, and journals slowly shift towards that direction. We perform multiclass evaluation to be aligned with the created dataset and baseline, but also present results with two settings to cater for multidisciplinarity. \topone where we output only the most-probable FoS Label and \toptwo where we output the two most-probable FoS Labels.
%%%%%%%%%%%%%%%%%%%%%%%%%%%%%%%%%%%%%%%%%%%%%%%%%%%%%%%%%%%%%%%%%%%
%%%%%%%%%%%%%%%%%%%%%%%%%%%%%%%%%%%%%%%%%%%%%%%%%%%%%%%%%%%%%%%%%%%

We compute \macrofone and \microfone to compare performance between \scinobo and \dann. Recall that our test set is unbalanced~(\ref{dataset}) and to account for it we also compute \weightedmacrofone.
%%%%%%%%%%%%%%%%%%%%%%%%%%%%%%%%%%%%%%%%%%%%%%%%%%%%%%%%%%%%%%%%%%%
%%%%%%%%%%%%%%%%%%%%%%%%%%%%%%%%%%%%%%%%%%%%%%%%%%%%%%%%%%%%%%%%%%%
\begin{table}[!t]
\resizebox{0.48\textwidth}{!}{
    \begin{tabular}{@{}llll@{}}
    \toprule
    \modelsresults        & \macrofone & \weightedmacrofone & \microfone \\ \midrule
    \danntopone            &0,4465 $\pm$ 0,003&	0,5873 $\pm$ 0,002&0,4563 $\pm$ 0,002\\
    \danntoptwo            &0,6007 $\pm$ 0,008&0,7307 $\pm$ 0,002&0,6154 $\pm$ 0,003\\
    \scinobocitreftopone &0,4627 $\pm$ 0,0&0,4900 $\pm$ 0,0&0,4900 $\pm$ 0,0\\
    \scinobocitreftoptwo &0.6000 $\pm$ 0,0&0,6208 $\pm$ 0,0&0,6177 $\pm$ 0,0\\
    \scinoboreftopone    &0,4781 $\pm$ 0,0&0,5000 $\pm$ 0,0&0,5022 $\pm$ 0,0\\
    \scinoboreftoptwo    &0,6190 $\pm$ 0,0&0,6277 $\pm$ 0,0&0,6205 $\pm$ 0,0\\
    \scinobopubtopone    &\textbf{0,7303 $\pm$ 0,0}&\textbf{0,7309 $\pm$ 0,0}&\textbf{0,7503 $\pm$ 0,0}\\
    \scinobopubtoptwo    &\textbf{0,8200 $\pm$ 0,0}&\textbf{0,8223 $\pm$ 0,0}&\textbf{0,8270 $\pm$ 0,0}\\ 
    \bottomrule
    \end{tabular}
}
\caption{\small{\macrofone, \weightedmacrofone and \microfone scores. Best scores are shown in bold. \dann's experiments were repeated 4 times and averaged results are presented with standard deviation. The postfix \topone and \toptwo refer to the two settings described in Section~\ref{fos_classification_evaluation}.}}
\label{tab:results}
\end{table}
\begin{figure*}[t]
\centering
\includegraphics[width=1.0\textwidth]{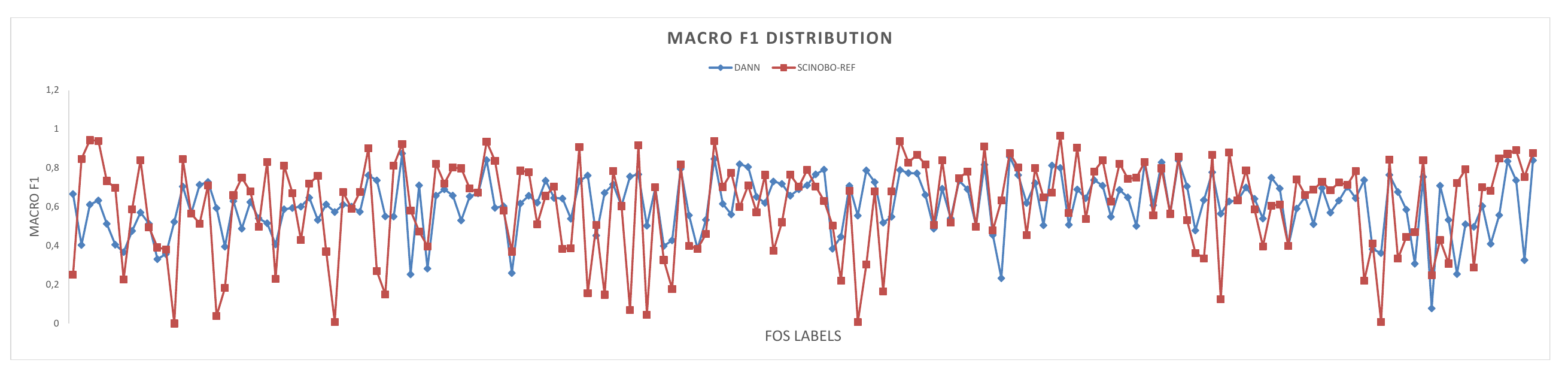}
\Description[Comparison between the DNN and SCINOBO-REF per field of science label] {Comparison showing that the MACRO-F1 score of SCINOBO-REF is better than that of DNN for the majority of FOS labels}
\caption{\macrofone distribution of all the FoS labels sorted by asceding order according to the number of instances. The names of the labels were omitted for simplicity.}
% \vspace*{50mm}

\label{f1_distribution}
\end{figure*}
\begin{table*}[!t]
\centering

    \resizebox{\textwidth}{!}{%
        \begin{tabular}{@{}lllllll@{}}
        \toprule
        \doi &
        \mytitle &
        % \myabstract &
        \publishedvenue &
        \scinoboref &
        \scinobocitref &
        \mylabel \\ \midrule
        10.3788/COL201715.062201 &
          \begin{tabular}[c]{@{}l@{}}Versatile nanosphere lithography technique combining multiple exposure\\ nanosphere lens lithography and nanosphere template lithography\end{tabular} 
          &
          Chinese Optics Letters &
          Optics &
          \begin{tabular}[c]{@{}l@{}}NanoScience \\ \& Technology\end{tabular} &
          Optics \\
          \midrule
          
        10.1089/ast.2019.2203 &
          \begin{tabular}[c]{@{}l@{}}The Role of Meteorite Impacts \\ in the Origin of Life\end{tabular}
          &
          Astrobiology &
          \begin{tabular}[c]{@{}l@{}}Astronomy\\ \& Astrophysics\end{tabular} &
          \begin{tabular}[c]{@{}l@{}}Developmental\\ Biology\end{tabular} &
          \begin{tabular}[c]{@{}l@{}}Astronomy\\ \& Astrophysics\end{tabular} \\
          \midrule
        10.1080/08839514.2018.1506971 &
          \begin{tabular}[c]{@{}l@{}}Strategic Particle Swarm Inertia \\ Selection for Electricity Markets\\ Participation Portfolio Optimization\end{tabular} 
          &
          \begin{tabular}[c]{@{}l@{}}Applied Artificial \\ Intelligence\end{tabular} &
          \begin{tabular}[c]{@{}l@{}}Artificial Intelligence\\ \& Image Processing\end{tabular} &
          Energy &
          \begin{tabular}[c]{@{}l@{}}Artificial Intelligence\\ \& Image Processing\end{tabular} \\ \bottomrule
        \end{tabular}
    }
\caption{\small{Publications presented with  metadata (\doi, \mytitle, \myabstract, \publishedvenue) along with the true FoS label of our dataset and the inferred FoS label of \scinoboref and \scinobocitref. Only snippets of the abstracts are presented.}}
\label{inference_examples}
\end{table*}
\normalfont
%%%%%%%%%%%%%%%%%%%%%%%%%%%%%%%%%%%%%%%%%%%%%%%%%%%%%%%%%%%%%%%%%%%
%%%%%%%%%%%%%%%%%%%%%%%%%%%%%%%%%%%%%%%%%%%%%%%%%%%%%%%%%%%%%%%%%%%
\subsection{FoS Classification Results}
\label{results}
%%%%%%%%%%%%%%%%%%%%%%%%%%%%%%%%%%%%%%%%%%%%%%%%%%%%%%%%%%%%%%%%%%%
%%%%%%%%%%%%%%%%%%%%%%%%%%%%%%%%%%%%%%%%%%%%%%%%%%%%%%%%%%%%%%%%%%%
Field of Science classification results are reported in Table.~\ref{tab:results}. Regarding \macrofone, \weightedmacrofone and \microfone our variant of \scinobo that utilizes only the published venues~(\scinobopub) outperforms all the other methods, in both evaluation settings (\topone, \toptwo). This presumably can be attributed to the nature of our evaluation dataset. Since the labelling of the publications in our dataset originates from labelling at the journal level, we can view \scinobopub as a method to perform journal classification, in effect same as \sciencemetrix. One would expect \scinobopub to perform much better than already did. This is not the case, because we re-evaluate the initial venue label assignment during Label Propagation~(\ref{propagation}). This label re-assignment originates from the neighborhood structure of the venue in question, indicating that the original label assignment should be reconsidered or that by aggregating more than one FoS labels the venue leaned towards multidisciplinarity.
%%%%%%%%%%%%%%%%%%%%%%%%%%%%%%%%%%%%%%%%%%%%%%%%%%%%%%%%%%%%%%%%%%%
%%%%%%%%%%%%%%%%%%%%%%%%%%%%%%%%%%%%%%%%%%%%%%%%%%%%%%%%%%%%%%%%%%%

\scinoboref clearly outperforms in \macrofone and \microfone both \scinobocitref and \dann in \topone setting and still outperforms them in \toptwo setting, however with the results being more competitive. We conclude that is important to take into account the references of a publication when it is published, which are always available (even on the first day of publication), unlike textual information. Furthermore, \dann's performance might be hindered by only utilizing abstracts, since many FoS labels have overlapping terminology~(\ref{motivation}). On the other hand, this effect is mitigated in \scinobo because authors usually cite similar work in their publications, making the FoS label more easily interpretable.
%%%%%%%%%%%%%%%%%%%%%%%%%%%%%%%%%%%%%%%%%%%%%%%%%%%%%%%%%%%%%%%%%%%
%%%%%%%%%%%%%%%%%%%%%%%%%%%%%%%%%%%%%%%%%%%%%%%%%%%%%%%%%%%%%%%%%%%

However, all of the methods perform much better in the \toptwo setting, revealing that the correct FoS label (according to the dataset) is frequently in the second most probable position. This implies the multidisciplinary nature of publications and suggests the need for creating a multi-label publication-level dataset to account for multidisciplinarity, which we leave for future work.
%%%%%%%%%%%%%%%%%%%%%%%%%%%%%%%%%%%%%%%%%%%%%%%%%%%%%%%%%%%%%%%%%%%
%%%%%%%%%%%%%%%%%%%%%%%%%%%%%%%%%%%%%%%%%%%%%%%%%%%%%%%%%%%%%%%%%%%

For \weightedmacrofone, \dann outperforms \scinobocitref and \scinoboref in both settings, achieving a high score in the \toptwo setting. Recall that our evaluation sets are unbalanced. \weightedmacrofone assigns a weight to the FoS labels according to the number of samples that each FoS label has in the evaluation set. \dann performs much better in this setting showing that it classifies correctly more high weighted FoS labels. Whereas our methods perform mostly similar in both metrics suggesting that presumably \scinobo overall generalizes better but performs poorly in some high-weighted FoS labels.
%%%%%%%%%%%%%%%%%%%%%%%%%%%%%%%%%%%%%%%%%%%%%%%%%%%%%%%%%%%%%%%%%%%
%%%%%%%%%%%%%%%%%%%%%%%%%%%%%%%%%%%%%%%%%%%%%%%%%%%%%%%%%%%%%%%%%%%

One key observation, is that \scinoboref performs slightly better than \scinobocitref in all three metrics and in both settings, suggesting that as time evolves and publications receive more and more citations their primary FoS label shifts thematically.
%%%%%%%%%%%%%%%%%%%%%%%%%%%%%%%%%%%%%%%%%%%%%%%%%%%%%%%%%%%%%%%%%%%
%%%%%%%%%%%%%%%%%%%%%%%%%%%%%%%%%%%%%%%%%%%%%%%%%%%%%%%%%%%%%%%%%%%

Finally, \dann is a deep neural attentive network and apart from computational training time (23 hours per experiment) with each experiment, results slightly deviate. Furthermore, differences in hardware, limit the reproducibility of the results. \scinobo has no computational/hardware overhead, apart from Label Propagation which is in the order of minutes and does not require a GPU, deviation of the results is non-existent which makes it more robust. The output will deviate only if the multilayer graph is changed or a publication receives more citations.
%%%%%%%%%%%%%%%%%%%%%%%%%%%%%%%%%%%%%%%%%%%%%%%%%%%%%%%%%%%%%%%%%%%
%%%%%%%%%%%%%%%%%%%%%%%%%%%%%%%%%%%%%%%%%%%%%%%%%%%%%%%%%%%%%%%%%%%
%%%%%%%%%%%%%%%%%%%%%%%%%%%%%%%%%%%%%%%%%%%%%%%%%%%%%%%%%%%%%%%%%%%
%%%%%%%%%%%%%%%%%%%%%%%%%%%%%%%%%%%%%%%%%%%%%%%%%%%%%%%%%%%%%%%%%%%
\subsection{Qualitative analysis}
\label{qualitative analysis}
To better understand the differences of each approach and to further establish the aforementioned arguments in Section~\ref{results}, we present qualitative results.
%%%%%%%%%%%%%%%%%%%%%%%%%%%%%%%%%%%%%%%%%%%%%%%%%%%%%%%%%%%%%%%%%%%
%%%%%%%%%%%%%%%%%%%%%%%%%%%%%%%%%%%%%%%%%%%%%%%%%%%%%%%%%%%%%%%%%%%

\noindent \textbf{Inferring with \scinoboref and \scinobocitref}: Table~\ref{inference_examples} presents three publications along with additional metadata (\doi, \mytitle, \myabstract) and the inferred FoS label when inferring with \scinoboref and \scinobocitref. We observe that \scinoboref inferred all three publications successfully. 

This behavior can be attributed to the fact that when a publication is published, it cites very similar content-wise publications. By examining the titles and abstracts, we can verify that the labelling of \scinoboref is correct. However, it is evident that all three publications can be applied to other FoS as well. For example, the third publication examines an application of artificial intelligence in energy. As time evolves, the publication received citations from other works that are involved in energy and are published in energy venues. This behavior shows the shift of FoS labels in publications with time and shows the importance of creating a publication-level multi-labeled dataset. The above-mentioned argument explains the lower \macrofone and \microfone scores in \scinobocitref, which according to the labelling of the dataset inferred the publications incorrectly.

%%%%%%%%%%%%%%%%%%%%%%%%%%%%%%%%%%%%%%%%%%%%%%%%%%%%%%%%%%%%%%%%%%%
%%%%%%%%%%%%%%%%%%%%%%%%%%%%%%%%%%%%%%%%%%%%%%%%%%%%%%%%%%%%%%%%%%%

\noindent \textbf{Case study for \scinobo and \dann}:
In Figure~\ref{f1_distribution}, we present the \macrofone distribution of \scinoboref and \dann for all the FoS labels, ordered by ascending order according to the number of instances. Recall that \dann only outperforms \scinobo and its variants in the \weightedmacrofone setting. Note that the \macrofone's scores remain the same in both settings but in \weightedmacrofone they are weighted according to the number of instances (support) each FoS label has in the evaluation set. We observe that the overall performance of \scinoboref is better than \dann. However, \scinoboref performs poorly in some of the FoS labels, whereas \dann performs fairly well in them indicating the reason for the high \weightedmacrofone score of \dann.
%%%%%%%%%%%%%%%%%%%%%%%%%%%%%%%%%%%%%%%%%%%%%%%%%%%%%%%%%%%%%%%%%%%
%%%%%%%%%%%%%%%%%%%%%%%%%%%%%%%%%%%%%%%%%%%%%%%%%%%%%%%%%%%%%%%%%%%
 \section{Conclusions and Future Work}
 \label{conclusions}
We propose \scinobo, a new method to perform Field of Science (FoS) classification along with a new taxonomy based on the classification scheme of the OECD disciplines/fields of research and development (FORD) and \sciencemetrix journal classification. \scinobo along with the FoS taxonomy are inherently hierarchical and can perform multi-label and multiclass evaluations accross all disciplines as opposed to previous work. Our proposed method leverages the strengths of utilizing minimal metadata that are always available, even at the first day a publication is published. By incorporating citing/publishing relationships into a Multilayer Graph containing publications-venues-FoS Labels, we are able to provide "context-aware" classifications without relying on the publication content as in many previous works. Furthermore, since our method can utilize citations that publications received, we can perform case studies showcasing the multidisciplinary nature of publications and how they can be assigned to more than one FoS labels in the course of time. We evaluated our method in a dataset created from \sciencemetrix classification and \acmag publications. Even though our dataset and baseline support only multiclass evaluation, experimental results and qualitative analysis demonstrated that our method is effective and outperforms a deep-learning method which rely solely on abstract information.

In future work, we plan to create a hierarchical publication-level multi-labeled dataset to better understand the benefits of \scinobo and to encourage further research. In addition, we plan to extend our FoS taxonomy to broader levels, to provide a better sense of granularity, which will also help us to identify emerging FoS labels and classify publications to them. Moreover, we will explore alternative data sources, such as the OpenAlex\footnote{\url{https://docs.openalex.org/}} public repository, as well as, alternative ways of assessing relations between publications beyond direct citation scores.

\begin{acks}
The authors thank the three anonymous reviewers for their constructive feedback. The authors would also like to thank Dimitris Pappas for his invaluable support on data collection, conceptualization of the layer-layer inference, and proposal of the multiplication of normalized weights during the inference step. This work  was supported by research grants from the \grantsponsor{1}{European Union’s Horizon 2020 Research and Innovation Programme}{} under grant agreement \grantnum[https://cordis.europa.eu/project/id/101004870]{1}{101004870} and the \grantsponsor{2}{EOSC Future Project}{} under grant agreement \grantnum[https://cordis.europa.eu/project/id/101017536]{2}{101017536}.
\end{acks}

\newpage

\bibliographystyle{ACM-Reference-Format}
\bibliography{references}
\end{document}